\documentclass[aps,superscriptaddress,amsmath,amssymb,twocolumn,showpacs,floatfix,english,noshowpacs]{revtex4}

\usepackage{url}
\usepackage{bm,bbm}
\usepackage{graphicx}
\usepackage{color}
\usepackage[colorlinks=true, urlcolor=blue, linkcolor=blue, citecolor=blue, pdftex]{hyperref}
\usepackage[english]{babel}

\usepackage{soul}
\usepackage{blindtext}
\usepackage{lipsum}

\newcommand{\dagga}{{\phantom{\dagger}}}

\begin{document}

\title{Effects of spin-phonon coupling in frustrated Heisenberg models}

\author{Francesco Ferrari}
\affiliation{Institute for Theoretical Physics, Goethe University Frankfurt, Max-von-Laue-Stra{\ss}e 1, D-60438 Frankfurt a.M., Germany}
\author{Roser Valent\'\i}
\affiliation{Institute for Theoretical Physics, Goethe University Frankfurt, Max-von-Laue-Stra{\ss}e 1, D-60438 Frankfurt a.M., Germany}
\author{Federico Becca}
\affiliation{Dipartimento di Fisica, Universit\`a di Trieste, Strada Costiera 11, I-34151 Trieste, Italy}

\date{\today}

\begin{abstract}
The existence and stability of spin-liquid phases represent a central topic in the field of frustrated magnetism. While a few examples of 
spin-liquid ground states are well established in specific models (e.g. the Kitaev model on the honeycomb lattice), recent investigations 
have suggested the possibility of their appearance in several Heisenberg-like models on frustrated lattices. An important related question 
concerns the stability of spin liquids in the presence of small perturbations in the Hamiltonian. In this respect, the magnetoelastic interaction 
between spins and phonons represents a relevant and physically motivated perturbation, which has been scarcely investigated so far. In this 
work, we study the effect of the spin-phonon coupling on prototypical models of frustrated magnetism. We adopt a variational framework based 
upon Gutzwiller-projected wave functions implemented with a spin-phonon Jastrow factor, providing a full quantum treatment of both spin and 
phonon degrees of freedom. The results on the frustrated $J_1-J_2$ Heisenberg model on one- and two-dimensional (square) lattices show that, 
while a valence-bond crystal is prone to lattice distortions, a gapless spin liquid is stable for small spin-phonon couplings. In view of 
the ubiquitous presence of lattice vibrations, our results are particularly important to demonstrate the possibility that gapless spin liquids 
may be realized in real materials.
\end{abstract}

\maketitle

\section{Introduction}

The physical properties of solid-state materials are ultimately governed by very simple physical laws, i.e., the Coulomb interaction among 
charged particles, electrons and nuclei. However, the low-energy physics of these many-body systems displays a variety of different behaviors, 
with emerging elementary and collective excitations, such as phonons, excitons, sound waves, magnons and Higgs modes, to mention a few.
This fact has been beautifully described by P.W. Anderson in his milestone paper ``More is different''~\cite{anderson1972}. Quantum spin 
liquids represent an amazing realization of this concept, since they exhibit long-range entanglement and absence of any local symmetry 
breaking~\cite{balents2010,savary2017}. Spin liquids can be divided into two broad classes, gapped and gapless, according to the presence or 
absence of a gap in the excitation spectrum. While the former ones are expected to be fully stable with respect to small perturbations, the 
latter ones are much more fragile, being inclined to develop some sort of symmetry breaking, such as valence-bond order~\cite{read1990}. More 
exotic instabilities have been also discussed, e.g., a topological phase with non-abelian anyonic excitations which is induced by magnetic 
fields in the Kitaev model on the honeycomb lattice ~\cite{kitaev2006}. 

One of the difficulties in detecting quantum spin liquids is the fact that the characteristic energy scale is given by the exchange coupling 
$J$, or even a small fraction of it, because of magnetic frustration. Therefore, small perturbations (e.g., disorder) may have strong
effects~\cite{riedl2019,dressel2021}. Phonons are also characterized by small energy scales (i.e., the Debye frequency $\omega$), with 
important effects on electronic properties. In particular, the super-exchange coupling between magnetic moments is affected by lattice 
distortions, since it depends upon the relative distance between the two ions where spins (electrons) are localized~\cite{fennie2006,zhang2008,lu2015}.
As a consequence, it may be profitable for the whole system (phonons and spins) to sacrifice some of the elastic energy in favor of the one 
gained by creating singlets, which optimize the magnetic energy of two spins~\cite{peierls1955}. For example, within an adiabatic approximation, 
where the kinetic energy of ions is neglected and lattice displacements are treated as classical variables, the one-dimensional spin-$1/2$ 
Heisenberg model is unstable with respect to a static dimerization; for the onset of this instability an infinitesimally small spin-phonon 
coupling is sufficient~\cite{cross1979}, since the energy gain for a distortion is linear in the displacement, while the loss due to the elastic 
energy is quadratic. The adiabatic limit of spin-phonon models has been studied in detail for a variety of cases~\cite{feiguin1997,garcia1997,augier1998,augier2000,becca2003,zhang2008}.

On the other hand, the full quantum problem, in which both spins and phonons are treated quantum mechanically, is considerably harder than the
adiabatic limit. It is worth noting that the full quantum description is relevant for most materials (e.g., CuGeO$_3$~\cite{lemmens2003}), whenever
the phonon frequency is of the same order of magnitude of $J$. From a computational perspective, one of the complications comes from the infinite
Hilbert space, which allows for an unbounded number of phonons on each lattice site. Therefore, numerical approaches like exact diagonalizations
or density-matrix renormalization group (DMRG) require a truncation of the Hilbert space~\cite{wellein1998,bursill1999,pearson2010}, e.g., fixing
a maximum number of phonons on each site. In addition, DMRG is limited to quasi-one-dimensional systems, since it needs an exponentially large
amount of resources in two or more spatial dimensions. An alternative approach, based upon a perturbative expansion and the definition of effective
spin models, may be also pursued~\cite{uhrig1998,weisse1999}, but the generic features of the model for $\omega \approx J$ cannot be captured.
Finally, quantum Monte Carlo methods~\cite{sandvik1999,weber2021} do not have limitations coming from the infinite Hilbert space of phonons, but
they are restricted to cases in which the Hamiltonian has no sign problem and this circumscribes their applicability. In this respect, the most
interesting and challenging problems in which frustrating interactions are present cannot be assessed, at least at low temperatures.

In spite of all these technical aspects, it would be desirable to include the lattice effects in spin models, for two main reasons. From a very 
general perspective, the first one comes from the desire to formulate a microscopic description that contains as many relevant ingredients as 
possible. In this regard, the role of disorder, Dzyaloshinskii-Moriya terms, or ring-exchange couplings have been discussed~\cite{lacroix2011}, 
but little effort has been spent to clarify the effect of the spin-phonon interactions. Still, lattice displacements may cause structural 
distortions and relevant modifications in the magnetic interactions. The typical example is the dimerization in quasi-one-dimensional 
systems~\cite{boucher1996}. Therefore, understanding the influence of phonons on the low-energy behavior of a quantum magnet is an important 
issue. The second reason is related to a particular aspect of the field, which is however of central importance. It deals with understanding 
the actual stability of spin-liquid phases in frustrated magnets~\cite{balents2010,savary2017}. In recent years, there have been several 
investigations addressing the possibility that a spin-liquid phase may be realized in an extended region of the phase diagram of frustrated 
spin models, one of the most notable example being the $S=1/2$ Heisenberg model on the kagome lattice that is relevant for 
Herbertsmithite~\cite{mendels2007,han2012,jeschke2013,norman2016}. At present, it is extremely important to clarify which kind of mechanisms 
may favor spin liquids and which ones disfavor them. For example, a fervent activity focuses on the role of spin-orbit coupling, which may 
enhance frustration by inducing microscopic interactions that explicitly break the $SU(2)$ spin symmetry. In analogy with the case of the Kitaev 
model~\cite{kitaev2006}, this could dramatically help the stabilization of spin liquids~\cite{iaconis2018,maksimov2019}. On the contrary, other 
kind of interactions may be highly detrimental for spin liquids, such as the spin-phonon coupling that could favor valence-bond crystals as it 
happens in one-dimensional systems. The question of the stability is particularly important for gapless spin liquids, which are considered to 
be more fragile to external perturbations. However, the analogy with the one-dimensional Heisenberg model with quantum phonons, where a finite 
critical value of the spin-phonon coupling is necessary to induce a spin-Peierls transition, may suggest that spin-liquids could be stable against 
moderate lattice distortions.

\begin{figure}
\includegraphics[width=0.85\columnwidth]{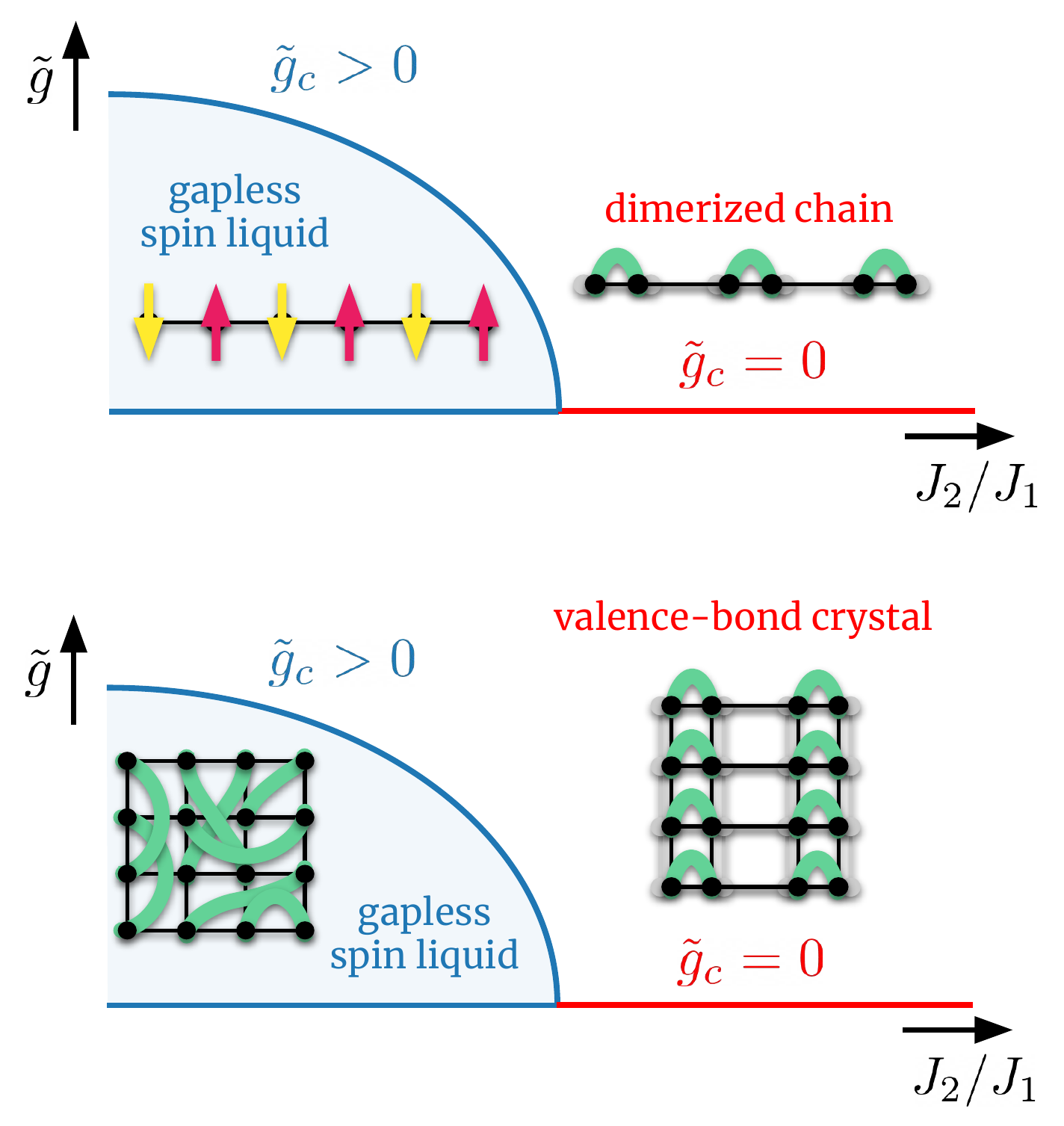}
\caption{\label{fig:phasediag}
Schematic illustration of the effect of the spin-phonon coupling on the $J_1-J_2$ model in one (upper panel) and two  dimensions (lower panel). 
In this work, we investigate how the gapless spin liquid phase in both models gets affected by the presence of spin-phonon coupling $\tilde{g}$.}
\end{figure}

In this work, we study the frustrated $J_1-J_2$ model in one and two dimensions, coupled to quantum phonons. We employ a variational Monte Carlo 
scheme, based upon a wave function that entangles spin and phonon degrees of freedom, which has been recently successfully benchmarked on an 
unfrustrated model~\cite{ferrari2020}. In one dimension, we report the critical line that separates the undistorted (gapless) quantum liquid 
from the distorted (gapped) spin-Peierls phase as a function of $J_2/J_1$, for two values of the phonon frequency $\omega$. The one-dimensional 
chain is prone to lattice distortions when the ground state of the pure spin model is gapped (i.e., for $J_2/J_1 \gtrsim 0.24$), while a finite 
spin-phonon coupling is necessary to open a spin gap and induce a distortion for $J_2/J_1 \lesssim 0.24$, i.e., where the pure spin model is 
gapless. The most important results are however for the $J_1-J_2$ model in two dimensions on the square lattice. In this case, recent studies 
suggested that the non-magnetic region in the proximity of $J_2/J_1 \approx 0.5$ consists of two different phases: a gapless spin liquid and a 
valence-bond crystal with columnar order~\cite{wang2018,ferrari2020b,nomura2020,liu2020}. When the coupling to quantum phonons is included, we 
find that the latter one is immediately unstable towards lattice distortions, as expected. By contrast, the gapless state remains stable for 
small spin-phonon couplings, supporting the fact that a gapless spin liquid may remarkably survive to magnetoelastic perturbations; however, when 
large enough spin-phonon couplings are considered, the same distortion of the valence-bond state (e.g., a columnar order of singlets) appears.
In Fig.~\ref{fig:phasediag}, we schematically display the effect of the spin-phonon coupling on the frustrated models under investigation.

The paper is organized as follows: in section~\ref{sec:method}, we describe the spin-phonon models in one and two dimensions and the variational 
wave functions; in section~\ref{sec:results}, we discuss the results; and finally, in section~\ref{sec:concl}, we draw our conclusions.

\section{Models and methods}\label{sec:method}

In this section, we present the spin-phonon models and the variational wave functions employed to obtain their ground-state properties. In order to 
make the presentation as clear as possible, we split the section in two parts: the first one deals with the one-dimensional case and the second one 
with the two-dimensional system.

\subsection{One-dimensional model}

In the one-dimensional $J_1-J_2$ model, $S=1/2$ spins (sitting on the sites of a linear chain) interact through antiferromagnetic first-neighbor 
($J_1>0$) and second-neighbor ($J_2>0$) Heisenberg exchange. We include magnetoelastic effects by assuming that the first-neighbor exchange is 
affected linearly by lattice distortions, analogously to what happens to the hopping terms of the Su-Schrieffer-Heeger (SSH) model~\cite{su1979}. 
Thus, the Hamiltonian of the SSH $J_1-J_2$ model is:
\begin{align}\label{eq:1dham}
 \mathcal{H}_{1d}&=J_1\sum_{r}  \left[1+ g (X_{r+1}-X_r) \right] \mathbf{S}_r \cdot \mathbf{S}_{r+1} \nonumber \\
 &+ J_2\sum_{r} \mathbf{S}_r \cdot \mathbf{S}_{r+2} +\frac{\omega}{4}\sum_{r} \left[P_{X,r}^{2} + X_r^{2}\right].
\end{align}
Taking the lattice spacing as $a=1$, we label the sites of the chain by their integer equilibrium positions $r=1,\dots,N$ and we consider periodic
boundary conditions. The ion mass $M$ has been absorbed in the definition of phonon displacements and their corresponding momenta, namely 
$X_r=\sqrt{2M\omega}x_r$ and $P_r=\sqrt{2/(M\omega)}p_r$, where $x_r$ and $p_r$ are the standard conjugate variables. Within this choice,
${P_{X,r}=-2i\frac{\partial}{\partial X_r}}$ and $[X_r,P_{X,r}]=2i \delta_{r,r'}$. We consider optical {\it Einstein phonons} with a flat 
dispersion and full quantum dynamics. The parameter $g$ measures the strength of the magnetoelastic coupling, while $\omega$ denotes the phonon
energy. For the sake of the upcoming discussion, we introduce the renormalized magnetoelastic parameter $\tilde{g}=(J_1/\omega)g$, which allows 
for an easier comparison of the results for different values of $\omega$.

We address the spin-phonon problem of Eq.~\eqref{eq:1dham} by a variational Monte Carlo approach. Our trial wave functions are products of a spin 
state ($\Psi_s$), a phonon condensate ($\Psi_p$) and a spin-phonon Jastrow factor ($\mathcal{J}_{sp}$):
\begin{equation}\label{eq:psi0}
 |\Psi_0\rangle=\mathcal{J}_{sp} |\Psi_s\rangle \otimes |\Psi_p\rangle.
\end{equation}
A detailed discussion of the variational method is given in Ref.~\cite{ferrari2020}, where a benchmark study on the Heisenberg model with quantum 
phonons [Eq.~\eqref{eq:1dham} with $J_2=0$] was performed. Here, we summarize the main features of the variational {\it Ans\"atze}.

The spin state $|\Psi_s\rangle$ is a Gutzwiller-projected fermionic state, whose definition stands on the Abrikosov fermion representation of
$S=1/2$ spins~\cite{abrikosov1965,savary2017}: the wave function is constructed by constraining a fermionic state, $|\Phi_0\rangle$, to the 
subspace of the fermionic Hilbert space in which each site is singly occupied. This operation, named {\it Gutzwiller projection}, yields a 
suitable state for spins and can be performed by an appropriate Monte Carlo sampling. Within our approach, the fermionic state $|\Phi_0\rangle$ 
to be projected is the ground state of an auxiliary BCS Hamiltonian 
\begin{equation}\label{eq:aux_H0}
{\cal H}_{0} = \sum_{r,r^\prime} \sum_{\sigma} t_{r,r^\prime} c_{r,\sigma}^\dagger c_{r^\prime,\sigma}^\dagga +
\sum_{r,r^\prime}  \Delta_{r,r^\prime} c_{r,\downarrow}^\dagga c_{r,\uparrow}^\dagga + h.c.,
\end{equation}
where $c_{r,\sigma}^\dagga$ and $c_{r,\sigma}^\dagger$ are the annihilation and creation operators of the Abrikosov fermion at site $r$ with spin 
$\sigma$. The Gutzwiller projection is represented by the operator ${{\cal P}_G = \prod_r n_r (2-n_r)}$, where 
${n_{r}=\sum_\sigma c^\dagger_{r,\sigma}c^\dagga_{r,\sigma}}$ is the local number operator. Thus, the full expression for $|\Psi_s\rangle$ reads
\begin{equation}
 |\Psi_s\rangle=\mathcal{J}_{ss} {\cal P}_G |\Phi_0\rangle,
\end{equation}
where, on top of the Gutzwiller-projected state, we have included also a long-range spin-spin Jastrow factor,
\begin{equation}
\mathcal{J}_{ss}=\exp\left[\sum_{r,r'} v_s(r,r') S_r^z S_{r'}^z\right]. 
\end{equation}
The variational parameters defining $|\Psi_s\rangle$ are the hopping ($t_{r,r^\prime}$) and pairing ($\Delta_{r,r^\prime}$) amplitudes of 
$\mathcal{H}_0$, and the pseudopotential parameters of the Jastrow factor, which are taken to be translationally invariant 
${v_s(r,r')=v_s\left(|r-r'|\right)}$. The second building block of the variational {\it Ansatz} of Eq.~\eqref{eq:psi0} is $|\Psi_p\rangle$, 
a phonon coherent state with momentum $k$. Its amplitude on a phonon configuration, labelled by the sites displacements $\{X_1,\dots,X_N\}$, 
is a product of Gaussian states
\begin{equation}
  \langle X_1,\dots,X_N|\Psi_p\rangle= \prod_{r=1}^N \exp[\phi_r(X_r)],
\end{equation}
where
\begin{equation}
 \phi_r(X_r) = iz\sin(k r)X_r-\frac{1}{4}[X_r-2z\cos(k r)]^2.
\end{equation}
To describe the Peierls distortion of the SSH $J_1-J_2$ chain, induced by the spin-phonon coupling, we take a coherent state with momentum 
$k=\pi$. We note that $z$ is another variational parameter, called {\it fugacity}, which controls the amplitude of the displacements in the 
phonon condensate~\cite{ferrari2020}. Finally, the last brick of our variational state is the spin-phonon Jastrow factor 
\begin{equation}\label{eq:jsp_1d}
 \mathcal{J}_{sp}=\exp\left[\frac{1}{2}\sum_{r,r'} v_{X}(r,r') S^z_r S^z_{r'}(X_r-X_{r'})\right],
\end{equation}
which entangles spin and lattice degrees of freedoms. The pseudopotential $v_{X}(r,r')$ depends only on the Euclidean distance between sites 
and is odd under the exchange of its arguments, namely ${v_{X}(r,r')=\tilde{v}_{X}(|r-r'|) \frac{r-r'}{|r-r'|}}$.

We finally remark that the variational wave function explicitly breaks the $SU(2)$ spin symmetry (due to the Jastrow factors, which are written
in terms of the $z$-component of the spin operators). This choice is dictated by the fact that the Monte Carlo sampling is performed within 
configurations with given spins along the $S^z$ axis and Jastrow factors containing also the other components of the spin operators would make
the numerical algorithm extremely more complicated. Nevertheless, the variational {\it Ansatz} is sufficiently accurate to obtain reliable 
results when compared to exact calculations on small systems~\cite{ferrari2020}.

\begin{figure}
\includegraphics[width=0.85\columnwidth]{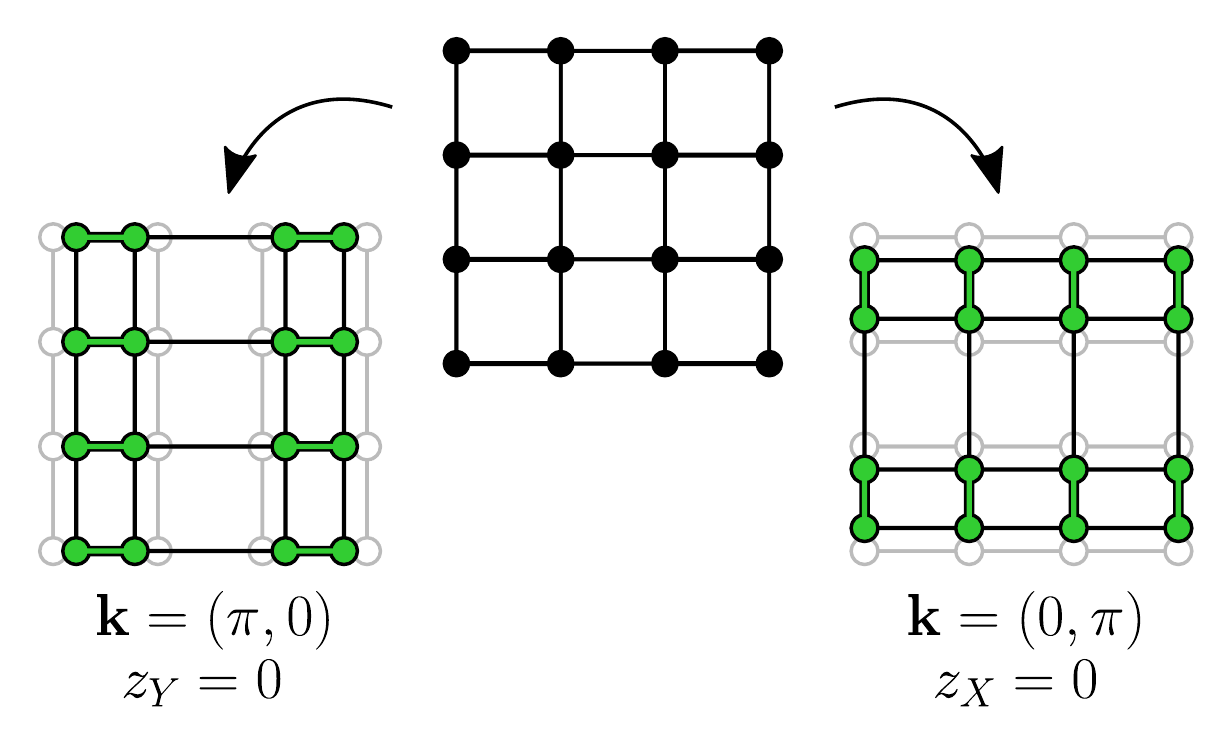}
\caption{\label{fig:distortions}
The two possible columnar lattice distortions of the square lattice SSH $J_1-J_2$ model. The one on the left involves only $X$ displacements, 
with momentum ${\bf k}=(\pi,0)$. The one on the right involves only $Y$ displacements, with momentum ${\bf k}=(0,\pi)$.}
\end{figure}

\subsection{Two-dimensional model}

In two dimensions, we study the generalization of the SSH $J_1-J_2$ model previously discussed. The spins of the system sit on the sites of a 
square lattice and interact through antiferromagnetic exchange at first- ($J_1$) and second-neighbors ($J_2$). The first-neighbor coupling is 
perturbed by lattice deformations in a SSH fashion:
\begin{align}\label{eq:2dham}
 \mathcal{H}_{2d}&=J_1\sum_{r}  \left[1+ g (X_{r+x}-X_r) \right] \mathbf{S}_r \cdot \mathbf{S}_{r+x} \nonumber \\
  &+J_1\sum_{r} \left[1+ g (Y_{r+y}-Y_r) \right] \mathbf{S}_r \cdot \mathbf{S}_{r+y}   \nonumber \\
 &+ J_2\sum_{r} \mathbf{S}_r \cdot \mathbf{S}_{r+x+y} + J_2\sum_{r} \mathbf{S}_r \cdot \mathbf{S}_{r+x-y} \nonumber \\
 &+\frac{\omega}{4}\sum_{r} \left[P_{X,r}^{2} +P_{Y,r}^{2} + X_r^{2}+ Y_r^{2}\right].
\end{align}
Here, ${\bf x}=(1,0)$ and ${\bf y}=(0,1)$ denote the unit vectors of the lattice. Periodic boundary conditions are adopted in both $x$ and $y$ 
directions. The system is characterized by two phonon modes per site ${\bf r}$, which are described by the operators $X_r$ and $Y_r$ measuring 
the displacements along the ${\bf x}$ and ${\bf y}$ directions, respectively; the corresponding momentum operators are 
${P_{X,r}=-2i\frac{\partial}{\partial X_r}}$ and ${P_{Y,r}=-2i\frac{\partial}{\partial Y_r}}$, thus leading to 
$[X_r,P_{X,r}]=[Y_r,P_{Y,r}]=2i\delta_{r,r'}$. As for the one-dimensional model, the phonons of the system are dispersionless, with frequency 
$\omega$, and their full quantum nature is taken into account. For simplicity, we assume that the magnetoelastic coupling affects only 
first-neighbor exchange: for sites ${\bf r}$ and ${\bf r}+{\bf x}$ (${\bf r}$ and ${\bf r}+{\bf y}$), the spin-spin interaction is linearly 
coupled to the difference of sites displacements along the direction of the bond, i.e., $X$-phonons ($Y$-phonons). The strength of the 
magnetoelastic coupling, denoted by $g$, is isotropic.

A generalization of the variational technique previously described is employed. The wave function has the same form as the one introduced in 
Eq.~\eqref{eq:psi0}, but its components are generalized to the two-dimensional case. The spin part, $|\Psi_s\rangle$, is a Gutzwiller-projected 
state defined on the square lattice, whose details are discussed in the next section. Concerning the phonon state $|\Psi_p\rangle$, we take the 
product of two coherent states, one for $X$-phonons and one for $Y$-phonons. As for the one-dimensional case, the amplitude of the state on a 
phonon configuration, specified by the displacements $\{X_r\}$ and $\{Y_r\}$ of each lattice site, is a product of Gaussian functions:
\begin{equation}
  \langle \{X_r\}; \{Y_r\}|\Psi_p\rangle= \prod_{r} \exp[\phi^X_r(X_r)] \exp[\phi^Y_r(Y_r)],
\end{equation}
where
\begin{align}
 \phi^X_r(X_r) &= iz_X\sin({\bf k}\cdot {\bf r})X_r-\frac{1}{4}[X_r-2z_X\cos({\bf k}\cdot {\bf r})]^2, \\
 \phi^Y_r(Y_r) &= iz_Y\sin({\bf k}\cdot {\bf r})Y_r-\frac{1}{4}[Y_r-2z_Y\cos({\bf k}\cdot {\bf r})]^2.
\end{align}
The phonon condensate is defined by the momentum ${\bf k}$ and two fugacity parameters, $z_X$ and $z_Y$, which can be optimized independently 
of each other. 

Finally, also the spin-phonon Jastrow factor is a generalization of the one-dimensional case and it consists of a product of two Jastrow terms, 
${\mathcal{J}_{sp}=\mathcal{J}^X_{sp}\mathcal{J}^Y_{sp}}$, one involving $X$-phonons
\begin{equation}\label{eq:jsp_2d_x}
 \mathcal{J}^X_{sp}=\exp\left[\frac{1}{2}\sum_{r,r'} v_{X}({\bf r},{\bf r}') S^z_r S^z_{r'}(X_r-X_{r'})  \right],
\end{equation}
and one involving $Y$-phonons
\begin{equation}\label{eq:jsp_2d_y}
 \mathcal{J}^Y_{sp}=\exp\left[\frac{1}{2}\sum_{r,r'} v_{Y}({\bf r},{\bf r}') S^z_r S^z_{r'}(Y_r-Y_{r'})  \right].
\end{equation}
The pseudopotentials $v_{X}$ and $v_{Y}$ fulfill the following properties
\begin{align}
 v_{X}({\bf r},{\bf r}')&=\tilde{v}_{X}(|{\bf r}-{\bf r}'|) \frac{({\bf r}-{\bf r}')}{|{\bf r}-{\bf r}'|}\cdot {\bf x}, \\
 v_{Y}({\bf r},{\bf r}')&=\tilde{v}_{Y}(|{\bf r}-{\bf r}'|) \frac{({\bf r}-{\bf r}')}{|{\bf r}-{\bf r}'|}\cdot {\bf y}.
\end{align}
The parameters $\tilde{v}_{X}$ and $\tilde{v}_{Y}$ are optimized for all the possible Euclidean distances on the square lattice.

The Peierls instability of the SSH $J_1-J_2$ model on the square lattice is associated with two equivalent columnar lattice distortions, depicted 
in Fig.~\ref{fig:distortions}. These distortions have a one-dimensional character, because they involve only one of the two phonon modes, either 
$X$ or $Y$. The columnar order along ${\bf x}$ corresponds to ${\bf k}=(\pi,0)$, $z_X \neq 0$ and $z_Y=0$ (no displacements along $Y$). On 
the contrary, the columnar order along ${\bf y}$ is described by ${\bf k}=(0,\pi)$, $z_Y \neq 0$ and $z_X=0$ (no displacements along $X$).
Alternative patterns for the dimerization have been also considered, limiting to the cases that may be relevant to the $J_1-J_2$ model without 
phonons. For example, a staggered valence-bond order can be described taking ${\bf k}=(\pi,\pi)$. In addition, a plaquette distortion can be 
obtained by slightly modifying the present formalism and introducing two different momenta for $\phi^X_r(X_r)$ and $\phi^Y_r(Y_r)$. Both 
the staggered and plaquette patterns do not give competitive variational energies compared to the columnar one. 
Therefore, in the following discussion, we focus on the the columnar Peierls distortions along ${\bf x}$, fixing the momenum ${\bf k}=(\pi,0)$.

As in one dimension, the variational state breaks the $SU(2)$ spin symmetry, due to the presence of Jastrow factors. We expect that also here
the accuracy of the wave function is sufficient to correctly reproduce the exact properties of the model. Unfortunately, direct comparisons with 
exact results are not numerically affordable, even in small two-dimensional clusters (e.g., $4 \times 4$).

\section{Results}\label{sec:results}

In this section, we discuss the numerical results of the SSH $J_1-J_2$ model in one and two dimensions.

\subsection{One-dimensional model}

The phase diagram of the spin-only $J_1-J_2$ Heisenberg model, without phonons, is well established: a critical point at $(J_2/J_1)_c \approx 0.24$ 
separates a gapless phase from a gapped one with dimer order. Extremely accurate estimates of the critical point have been achieved by level 
spectroscopy~\cite{eggert1996,sandvik2010}. In the unfrustrated limit $J_2/J_1=0$, the inclusion of quantum phonons as in Eq.~\eqref{eq:1dham} 
is known to drive a spin-Peierls transition, towards a gapped and dimerized phase, which is stabilized for a sufficiently large magnetoelastic 
coupling that depends on the frequency $\omega$~\cite{wellein1998,bursill1999,uhrig1998,sandvik1999}. Here, we presents our variational results 
for the generic case with both $J_1$ and $J_2$.

The auxiliary Hamiltonian~\eqref{eq:aux_H0} defining the spin part of the variational state contains both hopping and pairing terms. We start 
from the variational {\it Ans\"atze} employed in Ref.~\cite{ferrari2018} for the $J_1-J_2$ model without phonons, taking a fermionic Hamiltonian 
with hoppings at first- and third-neighbors, pairings at first- and second-neighbors, and a uniform onsite pairing. Then, we allow all first 
neighbor terms to break the translational symmetry of the lattice, assuming that they can be different on even and odd bonds. In the gapless 
phase, where the chain is undistorted, the translational symmetry is restored upon optimization of the variational energy. On the contrary, in 
the dimerized phase, the translational symmetry is explicitly broken by the first-neighbor terms of the fermionic Hamiltonian and by the 
displacements of the lattice sites, which form an alternating pattern of short and long bonds.

\begin{figure}
\includegraphics[width=0.85\columnwidth]{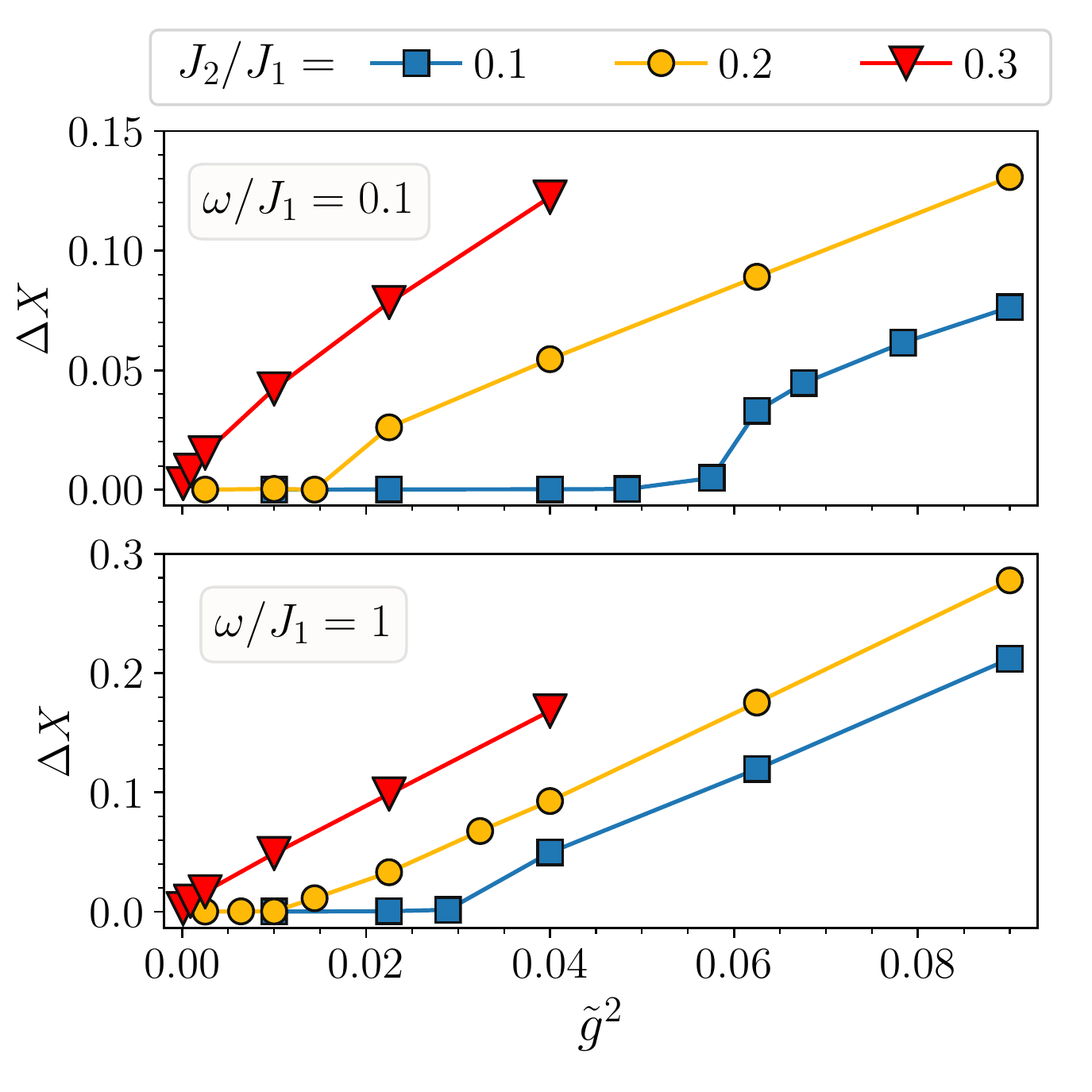}
\caption{\label{fig:1d_displ}
Distortion $\Delta X$ (Eq.~\ref{eq:deltax}) of the SSH $J_1-J_2$ Heisenberg chain as a function of $\tilde{g}^2$ with $\tilde{g}=(J_1/\omega)g$ 
for different values of the frustrating ratio $J_2/J_1$. The results have been obtained with a finite chain of $N=200$ sites. Upper and lower 
panels correspond to $\omega/J_1=0.1$ and $\omega/J_1=1$, respectively.}
\end{figure}

\begin{figure}
\includegraphics[width=0.9\columnwidth]{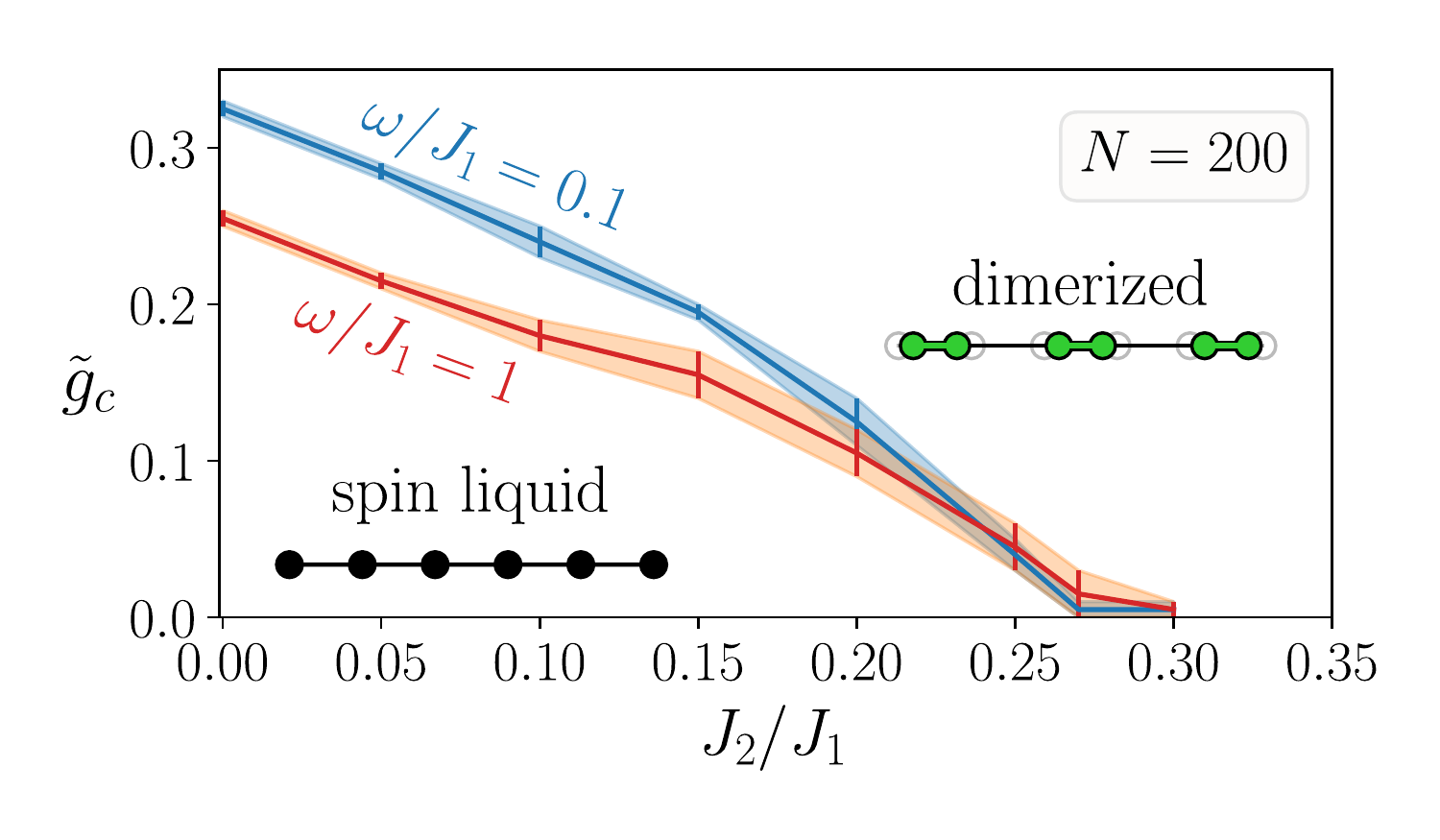}
\caption{\label{fig:phasediag_1d}
Phase diagram of the SSH $J_1-J_2$ Heisenberg chain, for $\omega/J_1=0.1$ and $\omega/J_1=1$. The results have been obtained with a finite chain 
of $N=200$ sites. The shaded areas denote the uncertainty on the estimates of the critical spin-phonon coupling $\tilde{g}_c$.}
\end{figure}

\begin{figure}
\includegraphics[width=\columnwidth]{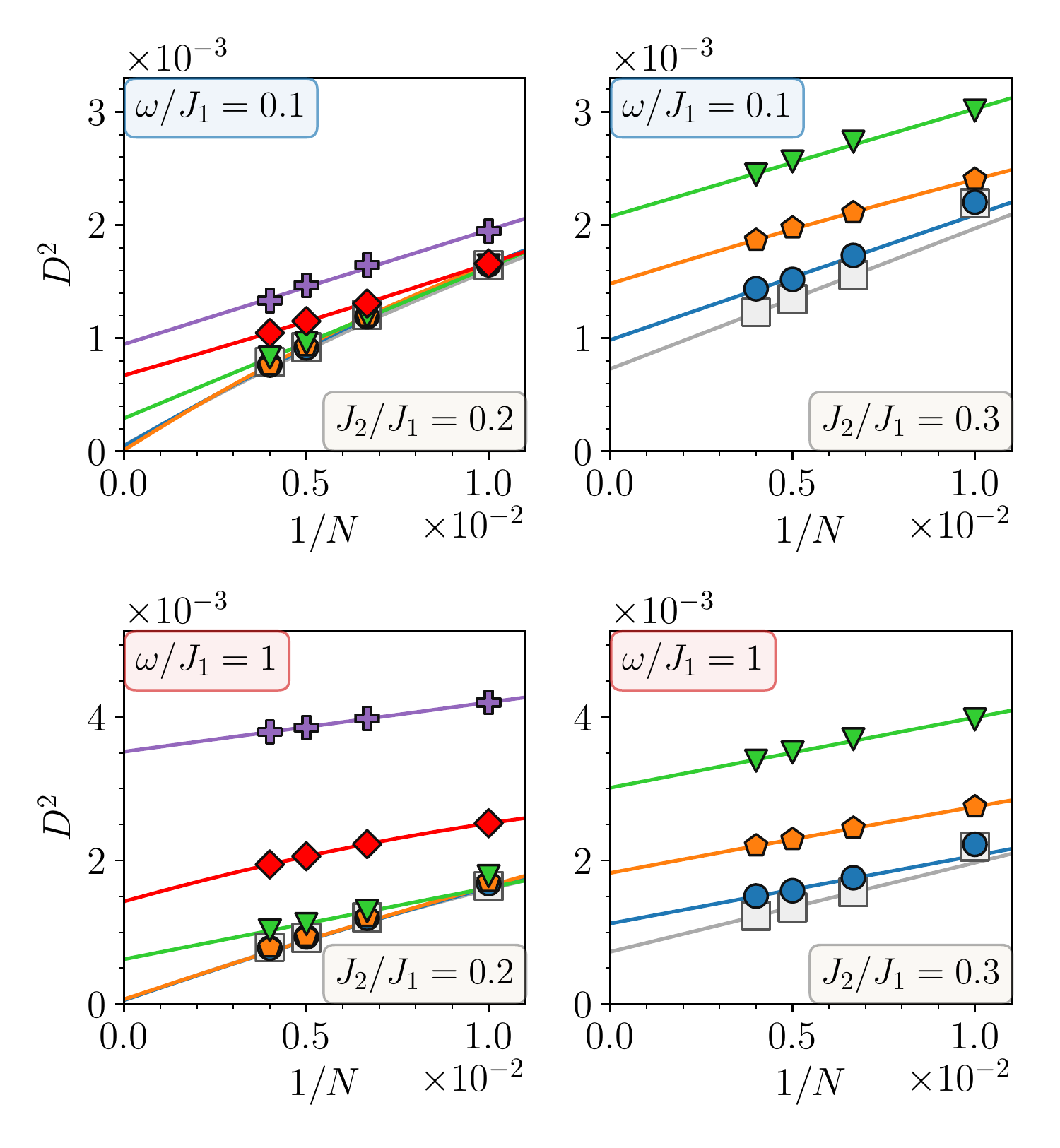}
\includegraphics[width=0.85\columnwidth]{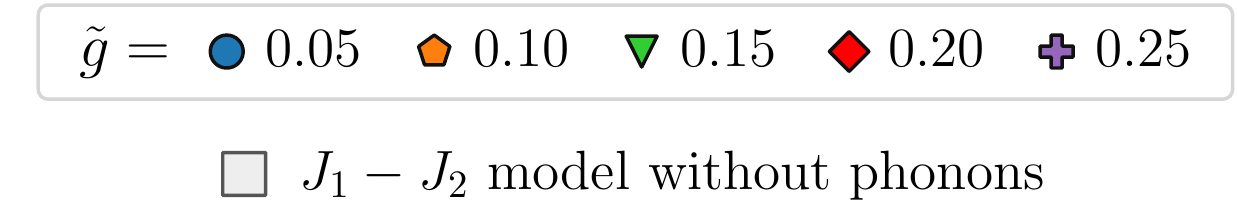}
\caption{\label{fig:1d_dimer}
Finite-size scaling of the dimer order parameter $D^2$ [Eq.~\eqref{eq:dpx}] of the SSH $J_1-J_2$ Heisenberg chain. Upper panels: results for 
$\omega/J_1=0.1$, $J_2/J_1=0.2$ (left panel) and $J_2/J_1=0.3$ (right panel). Lower panels: results for $\omega/J_1=1$, $J_2/J_1=0.2$ (left panel) 
and $J_2/J_1=0.3$ (right panel). The large grey squares represent the results obtained in absence of phonons, i.e., for the spin-only $J_1-J_2$ 
model.}
\end{figure}

The phase diagram of the SSH $J_1-J_2$ model can be obtained by assessing $k=\pi$ lattice distortions
\begin{equation}\label{eq:deltax}
 \Delta X=\left |\frac{1}{N}\sum_{r=1}^N e^{i\pi r} \langle X_r \rangle \right|, 
\end{equation}
and dimer-dimer correlations
\begin{equation}\label{eq:dpx}
  D^2=\frac{1}{N}\sum_{R=0}^{N-1} e^{i\pi R}  \left(\frac{1}{N} \sum_{r=1}^N \langle S^z_r S^z_{r+1} S^z_{r+R} S^z_{r+R+1} \rangle \right),
\end{equation}
where in both cases $\langle \dots \rangle$ stands for the expectation value over the variational wave function of Eq.~\eqref{eq:psi0}. In the 
gapless (undistorted) phase, both $D^2$ and $\Delta X$ vanish in the thermodynamic limit, while the gapped (distorted) phase is characterized 
by finite values of dimer-dimer correlations and lattice distortions. 

The behavior of $\Delta X$ for a large cluster with $N=200$ sites is shown in Fig.~\ref{fig:1d_displ}. Calculations are shown for three values 
of $J_2/J_1$, across the transition of the pure spin model: $J_2/J_1=0.1$ and $0.2$ are within the gapless phase, while ${J_2/J_1=0.3}$ is within 
the gapped phase. The results are obtained for two different values of the phonon frequency, $\omega/J_1=0.1$ and $1$. As in the unfrustrated 
Heisenberg model~\cite{ferrari2020}, a finite value of the spin-phonon coupling ${\tilde g}_c$ is needed to induce the lattice distortion in the 
gapless phase ($J_2/J_1=0.1$ and $0.2$); by contrast, in the gapped phase ($J_2/J_1=0.3$), $\Delta X$ is finite as soon as spins are coupled to 
phonons. Finite-size effects are small away from the critical point, especially when the phonon energy is not too small (e.g., they are smaller 
for $\omega/J_1=1$ than for $\omega/J_1=0.1$). Approaching the phase transition, they are more evident; for example, $\Delta X$ can be very small 
up to a given cluster size, becoming suddenly finite when the size exceeds a given value. Still, the critical point can be located with a 
sufficient precision. A systematic analysis of lattice displacements allows us to draw the phase diagram of Fig.~\ref{fig:phasediag_1d}. Here, 
the critical line, separating gapless and gapped phases, is compatible with the fact that ${\tilde g}_c$ goes to zero at $J_2/J_1 \approx 0.24$. 
Then, not surprisingly, when the pure spin model is already gapped and dimerized, an infinitesimal spin-phonon coupling is sufficient to generate 
a lattice distortion. 

A finite lattice distortion is always accompanied by spin dimerization, signalled by a finite value of $D^2$. However, while $\Delta X$ gives a 
rather sharp indication for the onset of distortions, dimer-dimer correlations require a more detailed size scaling. Indeed, on finite clusters, 
$D^2$ may be sizable also in the gapless region. In Fig.~\ref{fig:1d_dimer}, we show such analysis for $J_2/J_1=0.2$ and $0.3$. For the former 
case, $D^2$ scales to zero in the thermodynamic limit for small values of $\tilde{g}$, while it remains finite for sufficiently large values of 
the magnetoelastic coupling ($\tilde{g} \gtrsim 0.15$). Instead, for $J_2/J_1=0.3$, $D^2$ always extrapolates to a finite value in the thermodynamic
limit. 

\subsection{Two-dimensional model}

The two-dimensional $J_1-J_2$ model on the square lattice without phonons has been the subject of intensive investigations in the last 20 years, 
with contrasting results~\cite{sushkov2001,mambrini2006,richter2010,jiang2012,mezzacapo2012,wang2013,hu2013,gong2014,doretto2014,morita2015,poilblanc2017,haghshenas2018,liu2018,ferrari2018b,hering2019}.
The most debated (and interesting) region is in the vicinity of the highest-frustrated point $J_2/J_1=0.5$, where a magnetically disordered ground 
state should exist. Nevertheless, its physical properties have not been fully identified yet. Recently, some consensus is emerging, with evidences 
that two distinct phases may be present: a gapless spin liquid and a valence-bond solid~\cite{wang2018,ferrari2020b,nomura2020,liu2020}. Although 
there are small discrepancies in the precise location of transition points by different methods, variational calculations based upon 
Gutzwiller-projected fermionic wave functions suggested that a gapless ${\cal Z}_2$ spin liquid is stable for $0.48 \lesssim J_2/J_1 \lesssim 0.54$, 
while the valence-bond solid should be present for $0.54 \lesssim J_2/J_1 \lesssim 0.6$~\cite{ferrari2020b}. We remark that the existence of these 
two phases has been inferred from a level crossing between singlet and triplet states at low energy. By contrast, it is extremely difficult to 
assess the presence of valence-bond order directly from dimer-dimer correlations.

Let us now discuss the auxiliary Hamiltonian~\eqref{eq:aux_H0} defining the spin part of the variational state. This approach allows different 
spin-liquid {\it Ans\"atze}, which can be classified according to the so-called projective-symmetry group technique~\cite{wen2002}. Among them,
the best variational state of the $J_1-J_2$ model has an $s$-wave hoppings and $d$-wave pairings~\cite{hu2013}. The symmetry of the pairing terms
is $d_{x^2-y^2}$, for fermions on the opposite sublattice, or $d_{xy}$, for fermions on the same sublattice. The hopping is limited to first
neighbors, while the pairing includes first ($d_{x^2-y^2}$), second, and fifth neighbors ($d_{xy}$); within this {\it Ansatz}, the third-neighbor 
pairing is not allowed, while at forth neighbors a further $d_{x^2-y^2}$ would give a marginal energy improvement~\cite{hu2013}. After having 
selected the variational {\it Ansatz} for the phonon condensate, fixing the momentum ${\bf k}=(\pi,0)$ that determines the pattern of sites 
displacements, we allow the couplings in the auxiliary Hamiltonian to assume different values on shorter and longer bonds induced by the distortion.

\begin{figure}
\includegraphics[width=0.85\columnwidth]{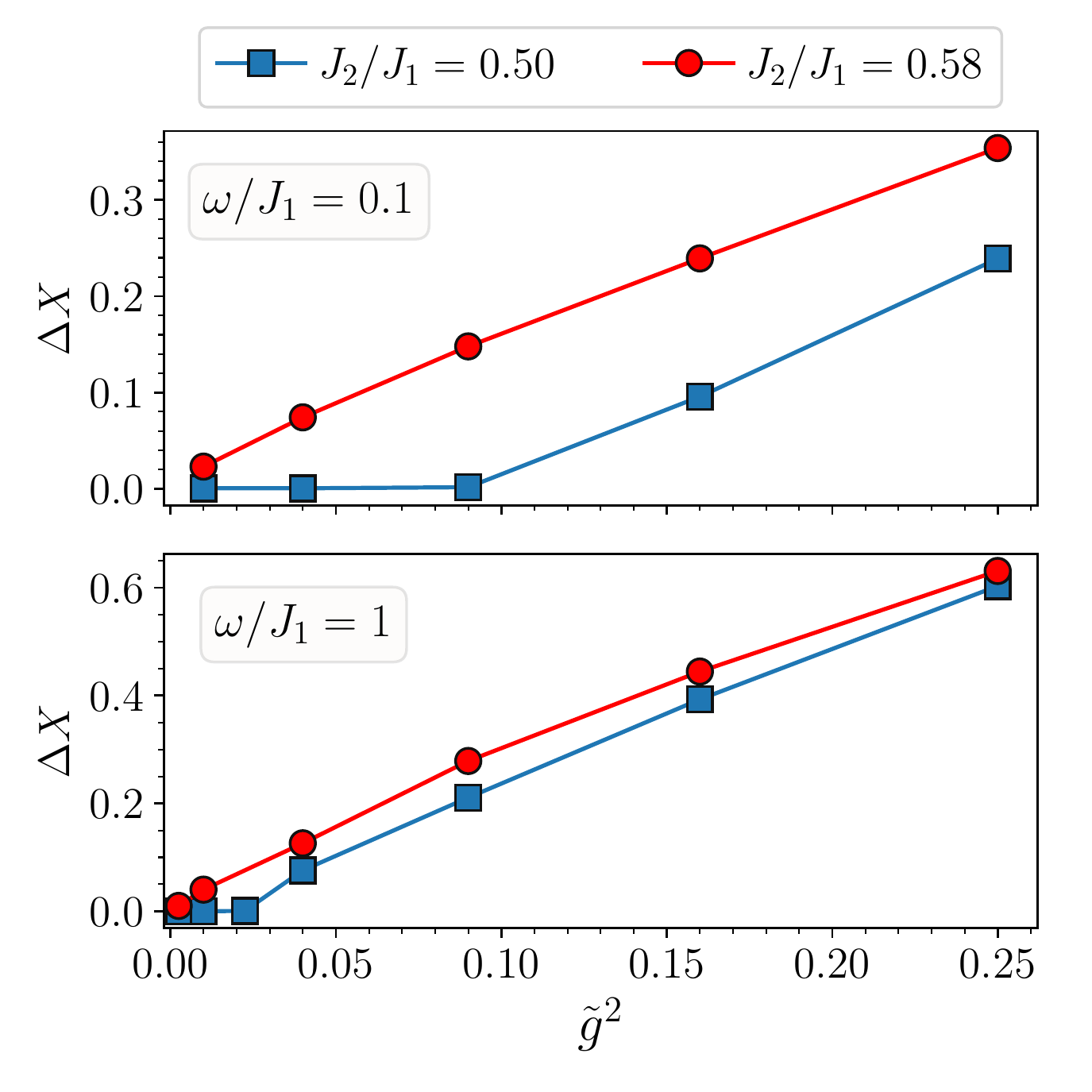}
\caption{\label{fig:2d_displ}
Lattice distortion order parameter $\Delta X$ [Eq.~\eqref{eq:delta2d}] for the SSH $J_1-J_2$ Heisenberg model on the square lattice as a function 
of $\tilde{g}^2$, for $J_2/J_1=0.50$ and $J_2/J_1=0.58$. The results have been obtained with a finite $16 \times 16$ cluster. Upper and lower 
panels correspond to $\omega/J_1=0.1$ and $\omega/J_1=1$, respectively.}
\end{figure}

\begin{figure}
\includegraphics[width=\columnwidth]{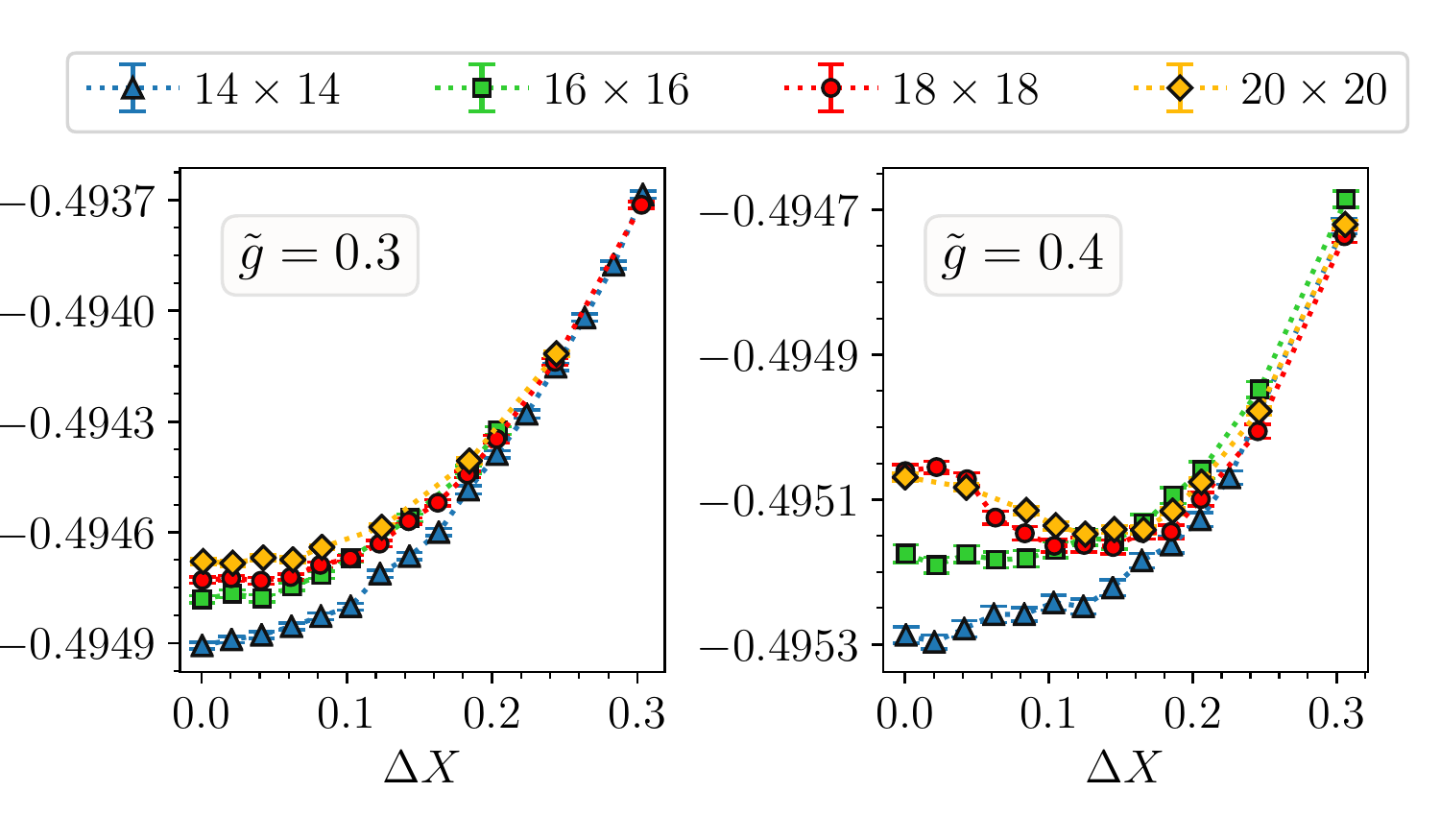}
\caption{\label{fig:landscape_size}
Variational energy as a function of the lattice distortion $\Delta X$ for the SSH $J_1-J_2$ Heisenberg model on the square lattice. The energy 
landscape has been computed for the case $\omega/J_1=0.1$ and $J_2/J_1=0.50$, $\tilde{g}=0.3$ (left panel) and $\tilde{g}=0.4$ (right panel). 
Finite clusters of different size have been used for the calculation ($L=14$, $16$, $18$, and $20$).}
\end{figure}

Having chosen ${\bf k}=(\pi,0)$ (leading to columnar order along ${\bf x}$), we define the distortion order parameter as
\begin{equation}\label{eq:delta2d}
 \Delta X=\left |\frac{1}{N}\sum_{r=1}^N e^{i {\bf k}\cdot {\bf r}} \langle X_r \rangle \right|,
\end{equation}
and the dimer-dimer correlations as
\begin{equation}\label{eq:dp2d}
  D^2=\frac{1}{N}\sum_{R} e^{i {\bf k}\cdot {\bf R}} \left(\frac{1}{N} 
\sum_{r} \langle S^z_{r} S^z_{r+x} S^z_{r+R} S^z_{r+R+x} \rangle \right).
\end{equation}
We focus on two values of the frustrating ratio, one in the gapless spin liquid phase ($J_2/J_1=0.50$) and the other in the valence-bond solid 
phase ($J_2/J_1=0.58$). In Fig.~\ref{fig:2d_displ}, we show the lattice distortion $\Delta X$ for $\omega/J_1=0.1$ and $1$, and different values 
of the spin-phonon coupling. The results are remarkably different for the two values of the frustrating ratio: while for $J_2/J_1=0.58$, the 
lattice is immediately distorted, i.e., as soon as an infinitesimal magnetoelastic coupling is included, for $J_2/J_1=0.50$ there are no 
appreciable distortions until the spin-phonon coupling reaches a finite critical value ${\tilde g}$. This observation suggests that the spin 
liquid is {\it stable} also in the presence of spin-lattice interactions. Given the gapless nature of the quantum spin liquid and its proximity 
to a valence-bond phase, this represents an absolutely non-trivial result.

As in one dimension, size effects are particularly relevant for small values of $\omega/J_1$ and in the vicinity of the phase transition, 
especially in the distorted regime. In fact, $\Delta X$ may be very small up to a given cluster size and then, abruptly, may become finite. 
In order to quantify size effects, in Fig.~\ref{fig:landscape_size} we show the landscape of the variational energy as a function of $\Delta X$, 
for different values of the cluster size. The different variational solutions forming the landscape are obtained by fixing the fugacity parameter 
$z_X$ to a set of different values and optimizing only the remaining parameters to get the lowest energy. Each of the solutions correspond to a 
state with a different value of $\Delta X$ and a different energy. In Fig.~\ref{fig:landscape_size} we concentrate on $J_2/J_1=0.50$ and we take 
two values of the magnetoelastic coupling (corresponding to undistorted and distorted cases). In the undistorted case (i.e., for ${\tilde g}=0.3$), 
size effects are under control, the landscape having a minimum at $\Delta X=0$, which is stable upon increasing $N$. By contrast, in the distorted 
case (i.e., for ${\tilde g}=0.4$), the landscape starts having a well defined minimum at $\Delta X \ne 0$ only for sufficiently large values of $N$. 
Still, the full Monte Carlo optimization (in which all parameters are optimized simultaneously) is able to detect the minimum at finite $\Delta X$ 
even when the landscape is very shallow, as demonstrated for the $16 \times 16$ cluster, where $\Delta X \approx 0.1$ is obtained, see 
Fig.~\ref{fig:2d_displ}. 

\begin{figure}
\includegraphics[width=\columnwidth]{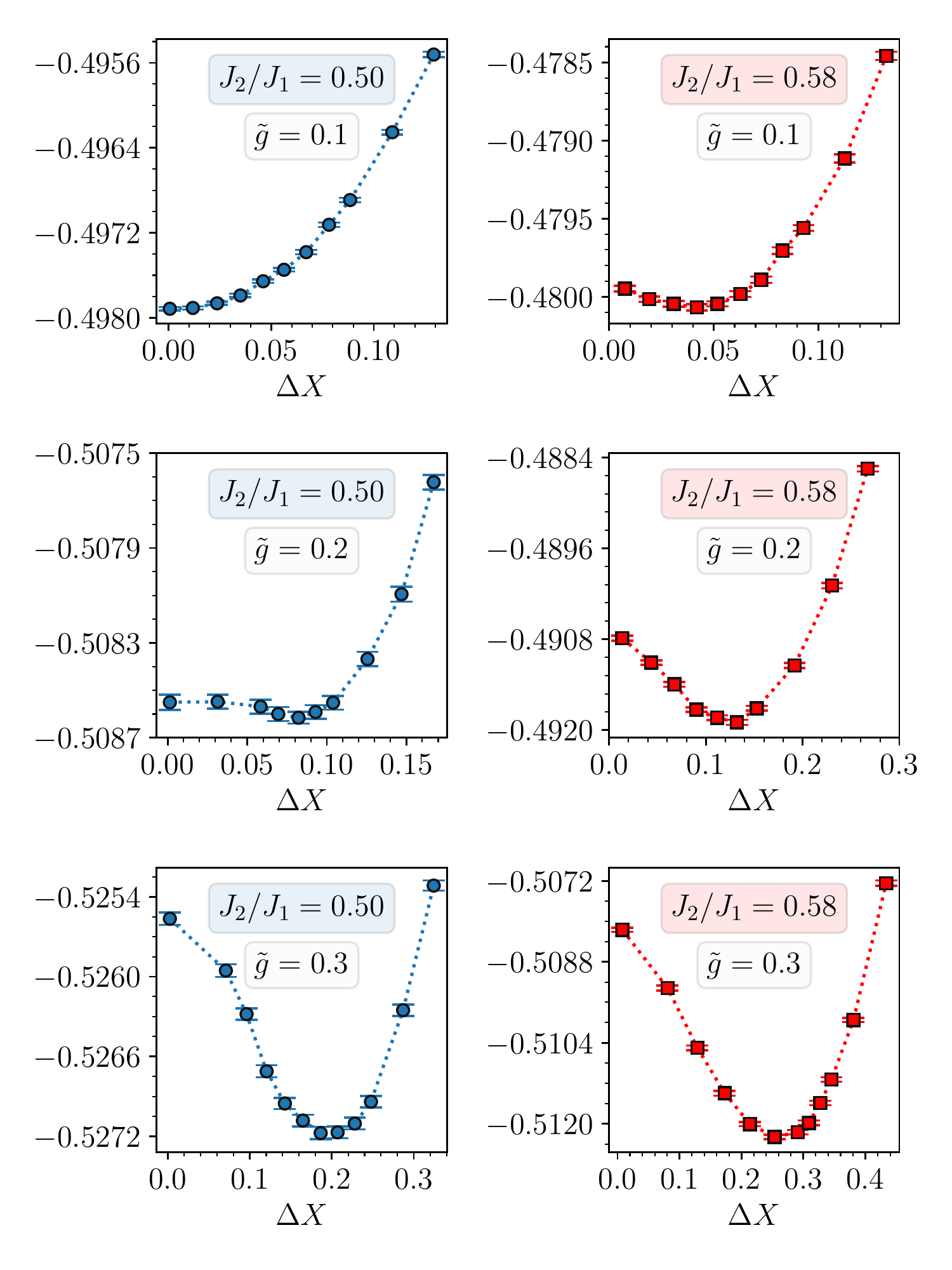}
\caption{\label{fig:landscape_omega1}
Variational energy as a function of the lattice distortion $\Delta X$ for the SSH $J_1-J_2$ Heisenberg model on the square lattice. The energy 
landscape has been computed for the case $\omega/J_1=1$, $J_2/J_1=0.50$ (on the left) and $J_2/J_1=0.58$ (on the right), and different values of 
$\tilde{g}$. The results have been obtained on a $16\times 16$ square lattice.}
\end{figure}

For $\omega/J_1=1$ size effects are much weaker than for $\omega/J_1=0.1$ and, therefore, relatively small clusters are sufficient to capture the
correct thermodynamic picture. In Fig.~\ref{fig:landscape_omega1}, we report the calculations of the energy landscape for the $16 \times 16$
cluster for both $J_2/J_1=0.50$ and $0.58$, for different values of ${\tilde g}$. The landscapes confirm our previous observation: a stable 
minimum at $\Delta X \ne 0$ is always present when including the spin-phonon coupling on the valence-bond solid ($J_2/J_1=0.58$); on the contrary, 
in the gapless spin-liquid regime ($J_2/J_1=0.50$), no distortions are visible if the magnetoelastic coupling is below a certain critical value 
$\tilde{g}_c$.

\begin{figure}
\includegraphics[width=\columnwidth]{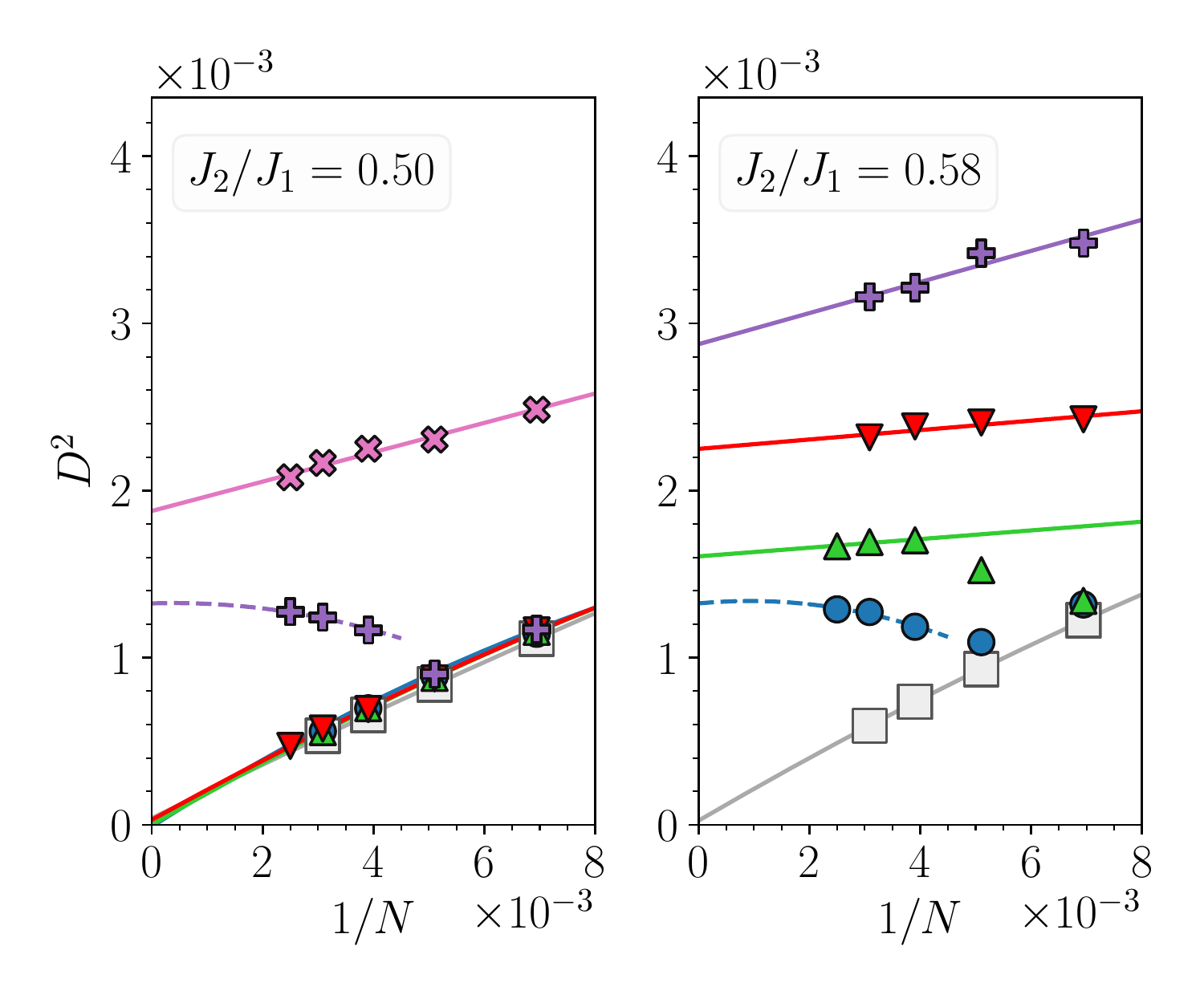}
\includegraphics[width=0.7\columnwidth]{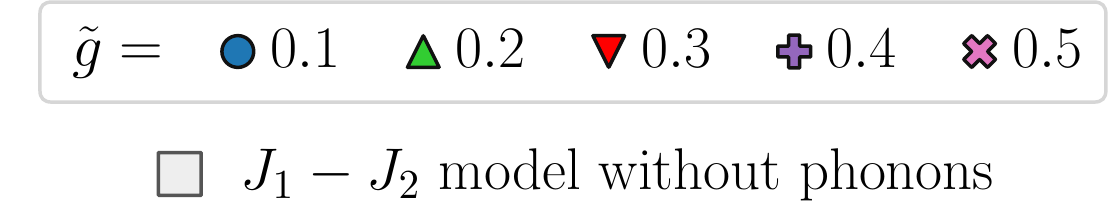}
\caption{\label{fig:2d_dimer_omega01}
Finite-size scaling of the dimer order parameter $D^2$ [Eq.~\eqref{eq:dp2d}]of the SSH $J_1-J_2$ Heisenberg model on the square lattice. Results 
for $\omega/J_1=0.1$ are shown: $J_2/J_1=0.50$ (left panel) and $J_2/J_1=0.58$ (right panel). The large grey squares represent the results obtained 
in absence of phonons, i.e., for the spin-only $J_1-J_2$ model.}
\end{figure}

\begin{figure}
\includegraphics[width=\columnwidth]{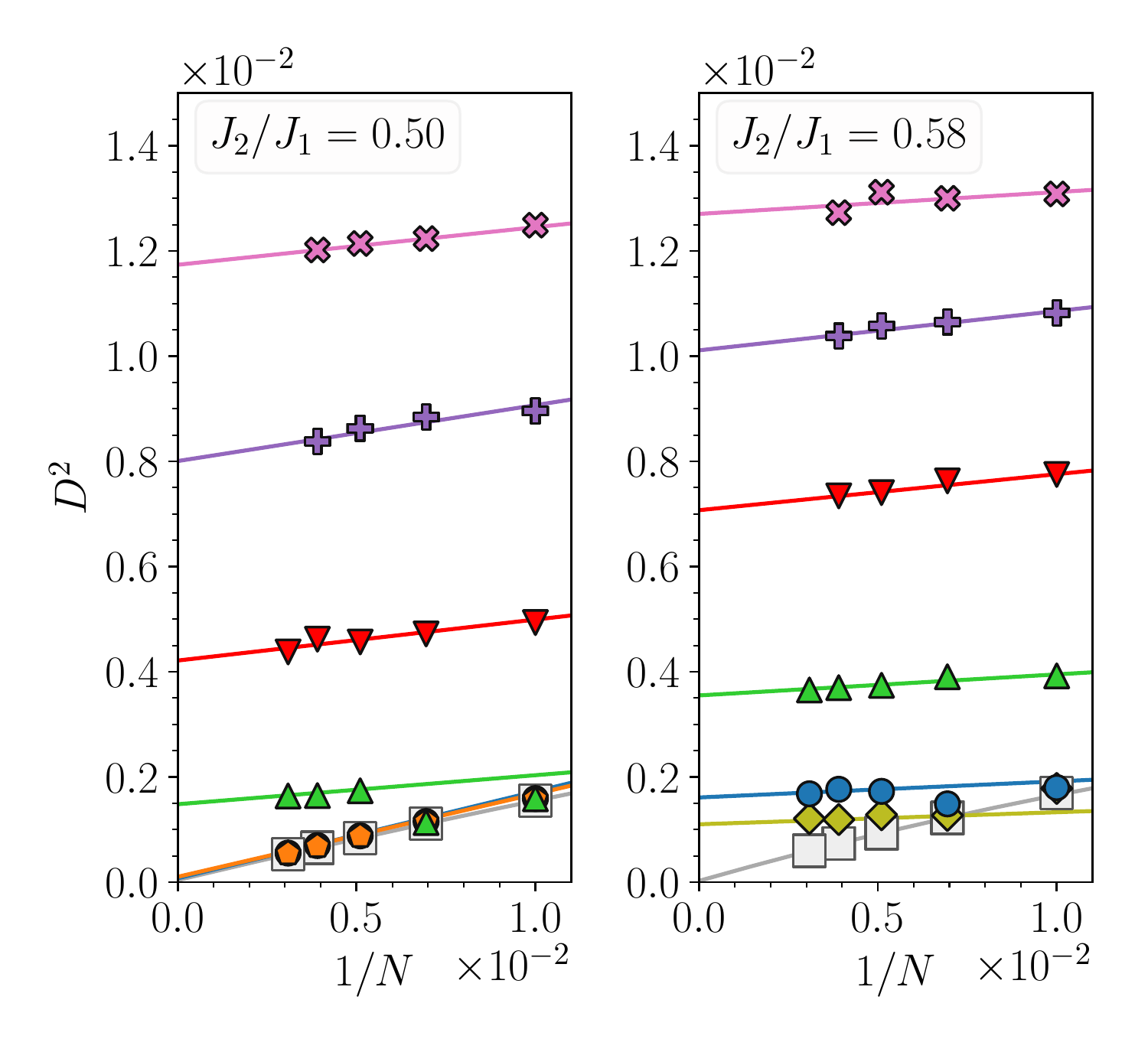}
\includegraphics[width=0.95\columnwidth]{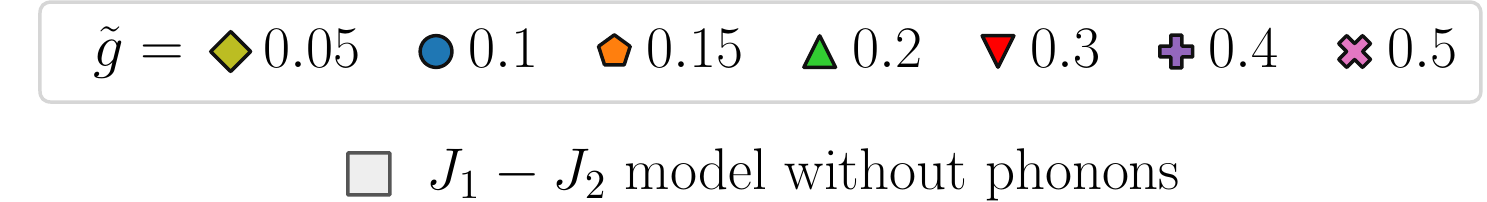}
\caption{\label{fig:2d_dimer_omega1}
The same as in Fig.~\ref{fig:2d_dimer_omega01} but for $\omega/J_1=1$.}
\end{figure}

Finally, we conclude our analysis with the dimer-dimer correlations, see Figs.~\ref{fig:2d_dimer_omega01} and~\ref{fig:2d_dimer_omega1}, where
the calculations for $\omega/J_1=0.1$ and $1$ are shown, respectively. The correlations follow the behavior of $\Delta X$, i.e.  $D^2$ scales to 
zero in the thermodynamic limit when $\Delta X \approx 0$, while a finite dimer order is detected in the cases in which $\Delta X \ne 0$. Notice 
that prominent size effects are visible for $\omega/J_1=0.1$ close to the transition, namely for ${\tilde g} \approx 0.4$ at $J_2/J_1=0.50$ and 
${\tilde g} \approx 0.1$ at $J_2/J_1=0.58$, where an accurate scaling is hard. Nevertheless, in both cases, the trend of the data indicate a finite 
value of $D^2$ in the thermodynamic limit. We note that in Figs.~\ref{fig:2d_dimer_omega01} and~\ref{fig:2d_dimer_omega1}, the values of $D^2$ of 
the pure spin model are also reported for comparison. As emphasized above, dimer-dimer correlations for the spin-only $J_1-J_2$ model on the square 
lattice do not show any appreciable long-range order in the thermodynamic limit, even in the putative valence-bond solid, i.e., at $J_2/J_1=0.58$.
However, it is remarkable that, in the presence of a magnetoelastic coupling, the spin-liquid and valence-bond phases react in a radically different 
way, with the latter immediately developing a finite value of $D^2$ in the thermodynamic limit.

\section{Conclusions}\label{sec:concl}

In this work, we investigated the effects of the spin-phonon coupling on two paradigmatic models of frustrated magnetism, namely the $J_1-J_2$ 
Heisenberg model in one and two dimensions (square lattice). Starting from the pure spin models, we introduced a magnetoelastic interaction by 
coupling the first-neighbor spin exchange to the relative displacements of lattice sites. The resulting models of spins and phonons have been 
tackled by a variational Monte Carlo approach, which incorporates the full quantum dynamics of the problem by means of Gutzwiller-projected 
fermionic states and Jastrow factors. The method does not suffer of sign problem in the presence of frustration, nor requires a truncation of 
the infinite Hilbert space of the phonons.

In the one-dimensional SSH $J_1-J_2$ model we track the evolution of the spin-Peierls transition, induced by the spin-phonon interaction, as a 
function of the frustrating exchange term ($J_2$). The onset of Peierls dimerization is assessed by measuring lattice distortions and dimer-dimer 
correlations. For small values of the frustrating ratio, a finite magnetoelastic coupling $\tilde{g}_c$ is necessary to drive the system from 
the gapless phase to the dimerized one, as in the simple SSH Heisenberg chain ($J_2=0$)~\cite{bursill1999}. However, the critical spin-phonon 
coupling of the Peierls transition decreases upon increasing $J_2/J_1$ and vanishes when the $J_1-J_2$ model enters the gapped phase with 
long-range dimer correlations ($J_2/J_1\gtrsim 0.24$). Here, an infinitesimally small magnetoelastic coupling is sufficient to induce a finite 
lattice distortion. 

The most interesting results are obtained for the $J_1-J_2$ model on the square lattice with spin-phonon interactions. We focused our attention 
on the highly-frustrated region of the pure spin model in the proximity of $J_2/J_1\approx 0.5$, whose nature is still undetermined. Recent 
numerical investigations indicated that the nonmagnetic region could be split into two phases, namely a gapless spin-liquid followed by a 
valence-bond solid~\cite{wang2018,ferrari2020b,nomura2020,liu2020}. At variance with the case of the one-dimensional $J_1-J_2$ model, the 
nonmagnetic region around the transition is extremely narrow and the dimer-dimer correlations do not differ appreciably in the two phases, 
thus making the presence of valence-bond order hardly detectable. However, the inclusion of magnetoelastic effects considerably enhances the 
difference between the two phases. For $J_2/J_1=0.58$, within the valence-bond ordered phase, the system develops a finite columnar distortion 
and long-range dimer order as soon as the spin-phonon coupling is included. On the contrary, for $J_2/J_1=0.50$, within the spin-liquid phase, 
a finite critical value of the magnetoelastic coupling is required to drive the system towards dimerization. This important result suggests that 
gapless spin liquids, which are believed to be fragile to external perturbations, could be stable with respect to the interaction between spins 
and lattice distortions.

The present results open the way to a number of applications of the variational method to other lattices (e.g., triangular and kagome) and 
more realistic frustrated spin models. In addition, it would be interesting also to study the effect of the spin-phonon coupling in magnetically
ordered states (especially for large values of $J_2/J_1$ where different spin-spin correlations are present along $x$ and $y$ directions) or in 
fully gapped spin liquids. Finally, the stability of spin liquid phases to other kinds of phonons, with an acoustic dispersion, represents a 
possible direction of future investigations.

\section*{Acknowledgments}
F.F. acknowledges support from the Alexander von Humboldt Foundation through a postdoctoral Humboldt fellowship. R.V. acknowledges the Deutsche 
Forschungsgemeinschaft (DFG, German Research Foundation) for funding through TRR 288 - 422213477 (project A05).

\end{document}